\documentstyle[12pt,epsf]{article}
\newcommand{\goes}{\rightarrow}
\def\lsim{\hbox{$\hskip0.3em\raisebox{-.4ex}{$\sim$}\raisebox{.4ex}{\hskip-0.8em
$<$\hskip0.1em}$}}

\newfont{\lfont}{line10}

\newcommand{\NN}{\nonumber}

\newcommand{\BE}{\begin{equation}}
\newcommand{\EE}{\end{equation}}
\newcommand{\BEA}{\begin{eqnarray}}
\newcommand{\EEA}{\end{eqnarray}}

\newcommand{\HEPTH}[1]{{\tt hep-th/{#1}}}
\newcommand{\IJMP}[3]{Int. J. Mod. Phys. {\bf #1}{(#2)}{#3}}
\newcommand{\JHEP}[3]{JHEP~{\bf #1}{(#2)}{#3}}

\newcommand{\NP}[3]{Nucl. Phys. {\bf #1}{(#2)}{#3}}
\newcommand{\PL}[3]{Phys. Lett. {\bf #1}{(#2)}{#3}}
\newcommand{\PR}[3]{Phys. Rev. {\bf #1}{(#2)}{#3}}
\newcommand{\PRL}[3]{Phys. Rev. Lett. {\bf #1}{(#2)}{#3}}

\def\ap{\alpha^{\prime}}
\def\p{\partial}

\def\12{\frac{1}{2}}
\def\bea{\begin{eqnarray}}
\def\eea{\end{eqnarray}}
\def\ba{\begin{array}}
\def\ea{\end{array}}

\def\one-loop{\mbox{\scriptsize one-loop}}

\def\G{\Gamma}

\catcode`\@=11

\def\@normalsize{\@setsize\normalsize{15pt}\xiipt\@xiipt
\abovedisplayskip 14pt plus3pt minus3pt%
\belowdisplayskip \abovedisplayskip
\abovedisplayshortskip  \z@ plus3pt%
\belowdisplayshortskip  7pt plus3.5pt minus0pt}

\def\small{\@setsize\small{13.6pt}\xipt\@xipt
\abovedisplayskip 13pt plus3pt minus3pt%
\belowdisplayskip \abovedisplayskip
\abovedisplayshortskip  \z@ plus3pt%
\belowdisplayshortskip  7pt plus3.5pt minus0pt
\def\@listi{\parsep 4.5pt plus 2pt minus 1pt
            \itemsep \parsep
            \topsep 9pt plus 3pt minus 3pt}}

\def\underline#1{\relax\ifmmode\@@underline#1\else
        $\@@underline{\hbox{#1}}$\relax\fi}
\@twosidetrue

\relax

\catcode`@=12

\catcode`\@=11

\def\section{\@startsection{section}{1}{\z@}{3.5ex plus 1ex minus
   .2ex}{2.3ex plus .2ex}{\large\bf}}

\def\thesection{\Roman{section}.}

\def\appendix{\setcounter{section}{0}
        \def\thesection{Appendix }
        \def\theequation{\Alph{section}.\arabic{equation}}}

\def\ps@headings{\def\@oddfoot{}\def\@evenfoot{}
\def\@oddhead{\hbox{}\hfill
        \makebox[.5\textwidth]{\raggedright\ignorespaces --\thepage{}--
        \hfill {}}}
\def\@oddhead{\hbox{}\hfill --\thepage{}-- \hfill
        {}}
\def\@evenhead{\@oddhead}
\def\subsectionmark##1{\markboth{##1}{}}
}

\ps@headings

\catcode`\@=12

\relax

\def\figcap{\section*{Figure Captions\markboth
        {FIGURECAPTIONS}{FIGURECAPTIONS}}\list
        {Fig. \arabic{enumi}:\hfill}{\settowidth\labelwidth{Fig. 999:}
        \leftmargin\labelwidth
        \advance\leftmargin\labelsep\usecounter{enumi}}}
 \relax
\def\tablecap{\section*{Table Captions\markboth
        {TABLECAPTIONS}{TABLECAPTIONS}}\list
        {Table \arabic{enumi}:\hfill}{\settowidth\labelwidth{Table 999:}
        \leftmargin\labelwidth
        \advance\leftmargin\labelsep\usecounter{enumi}}}
 \relax
\def\reflist{\section*{References\markboth
        {REFLIST}{REFLIST}}\list
        {[\arabic{enumi}]\hfill}{\settowidth\labelwidth{[999]}
        \leftmargin\labelwidth
        \advance\leftmargin\labelsep\usecounter{enumi}}}
 \relax

\catcode`\@=11

\def\ps@headings{\def\@oddfoot{}\def\@evenfoot{}
\def\@oddhead{\hbox{}\hfill
        \makebox[.5\textwidth]{\raggedright\ignorespaces --\thepage{}--
        \hfill {}}}
\def\@evenhead{\@oddhead}
\def\subsectionmark##1{\markboth{##1}{}}
}

\ps@headings

\relax

\newskip\humongous \humongous=0pt plus 1000pt minus 1000pt

\newif\ifdtup

\def\beq{\begin{equation}}
\def\eeq{\end{equation}}

\def\beqn{\begin{eqnarray}}
\def\eeqn{\end{eqnarray}}
\relax

\def\G2{{\; \rm GeV/}c^2}
\def\G{\; \rm GeV}

\def\dotx{\dotx{\dot\overline{x}}}

\relax

\def\p{\partial}

\begin{document}
%
%
\begin{titlepage}

\renewcommand{\thefootnote}{\fnsymbol{footnote}}

\begin{flushright}
      \normalsize
     December, 2000  \\     
     OU-HET 372, \\ 
         hep-th/0012150  \\
\end{flushright}

%
\begin{center}
  {\large\bf Behavior of Boundary String Field Theory \\
Associated with Integrable Massless Flow
}%
\end{center}

\vfill

\begin{center}
    {%
A.~Fujii\footnote{e-mail: fujii@het.phys.sci.osaka-u.ac.jp}
\quad and\quad 
H.~Itoyama\footnote{e-mail: itoyama@funpth.phys.sci.osaka-u.ac.jp}
}\\
\end{center}

\vfill

\begin{center}
      \it  Department of Physics,
        Graduate School of Science, Osaka University,\\
        Toyonaka, Osaka 560-0043, Japan
\end{center}

\vfill


\begin{abstract}
We put forward an idea that the boundary entropy associated with 
integrable massless flow of thermodynamic Bethe ansatz (TBA) is 
identified with tachyon action of boundary string field 
theory. We show that the temperature parameterizing a  
massless flow in the TBA formalism can be identified with  
tachyon energy for the classical action at least near the 
ultraviolet fixed point, {\it i.e.} the open string vacuum. 
\end{abstract}

\vfill

\setcounter{footnote}{0}
\renewcommand{\thefootnote}{\arabic{footnote}}

\end{titlepage}
A lot of efforts to reveal tachyon condensation mechanism 
have been made in an attempt to find a stable vacuum both in 
bosonic and in supersymmetric string theory. According to Sen's 
conjecture, the closed string vacuum is realized 
after an annihilation mechanism of an open string is completed by 
the cancellation between tensions of D-branes and energy of the 
tachyon\cite{sen}. 
In string field theory,  
main ways to analyze this problem have been through Witten's 
cubic string field theory\cite{wittencubic} 
and the boundary string field theory (BSFT)\cite{wittenbsft,shatashvili}. 

In the latter context with a special choice of the tachyon profile, 
some evidence to support Sen's conjecture has recently been  
provided in an exact manner
\cite{harvey-kutasov-Martinec}-\cite{kutasov-marino-moore}. 
In this letter we take this latter 
approach, BSFT. In the Batalin-Vilkovisky formalism used for  
BSFT\cite{wittenbsft}, 
the space-time string action $S$ is conjectured to 
satisfy the differential equation of the form
\cite{harvey-kutasov-Martinec}-\cite{kutasov-marino-moore}
\BE
{\p S\over\p\lambda_i }={\cal G}^{ij}\,\beta_j.
\label{eq:deS}
\EE
Here $\lambda_i$ are the coupling constants 
of the boundary operators, $\beta_j$ are the $\beta$-functions,
 and ${\cal G}^{ij}$ is  
the metric in the space of coupling constants. 

On the other hand, ground state degeneracy ($g$-function)
\cite{affleck-ludwig} of the worldsheet theory with a boundary 
perturbation  is also expected to satisfy a differential equation 
of the same form as (\ref{eq:deS}).  
Thus, we expect that string action $S$ will be identified 
with $g$-function. To investigate this 
correspondence, we analyze this problem using 
the boundary sine-Gordon model (BSG)\cite{bcft}  
and attendant thermodynamic Bethe ansatz (TBA).   
TBA is intended to obtain thermodynamic quantities 
at finite temperature and
has also been used to extract information on the   
$g$-function in some of exactly solvable models
\cite{fendley-saleur-warner}. 

We consider the process in which single D25-brane decays  
into a D24-brane by tachyon condensation.  
In the context of TBA, temperature in one-dimensional 
soliton gas is the renormalization group (RG) 
scale, which is regarded as an order parameter 
of the tachyon condensation. 
The sine-Gordon parameter in the boundary term 
can be identified with the inverse of the 
compactification radius in BSFT. 
From this fact, the boundary entropies at two ends of the 
flow have been shown to give an exact ratio of the brane 
tensions\cite{harvey-kutasov-Martinec}. 
In this letter we find that the temperature in TBA 
can be identified with energy of the classical solution 
of the tachyon action in BSFT. 
We provide evidence in support of 
this correspondence by comparing the behavior of TBA  
and the behavior of the classical solution in BSFT. 
This correspondence is confirmed by a numerical calculation, too.  
We propose that integrable massless flows generated by TBA 
provide description of the open string action even 
away from the fixed points. 
\newpage
{\bf Massless TBA and $g$-function as string action}

To utilize soliton picture, we begin with the action of the 
sine-Gordon 
model on a segment $\sigma\in \left[ 0,L\right]$ at finite 
temperature $\theta$
\BE
{\cal S}={1\over 4\pi\ap}\int^{1/\theta}_0 dt\int^{L}_{0}d\sigma
\left[ (\p_\mu X(\sigma,t))^2 +
G \cos{4\pi\over R}X(\sigma,t)\right] 
+\zeta\int^{1/\theta}_0 dt\cos{2\pi\over R}
X(\sigma=0,t).\label{eq:action}
\EE
This system 
is shown to possess an infinite number of conserved currents  
and hence is integrable\cite{bsG}. 
This action permits the field $X(\sigma,t)$, namely $X_{25}$,  
to be compactified: $X\sim X+R$. 
We impose the Dirichlet boundary condition at $\sigma=L$ 
and pay our attention to the boundary at $\sigma=0$. We assume  
that $\lambda=R^2 /4\pi^2\ap -1$ is a non-negative integer. 
The strength $G$ gives mass scale of the soliton/antisoliton 
and $\zeta$ gets traded with boundary temperature $\theta_B$
\cite{fendley-saleur-warner}, which plays a similar role to 
that of the Kondo temperature in the Kondo problem.  
Because we are interested in models with conformal invariance 
in the bulk $\sigma\in (0,L)$, 
the massless limit $G\goes 0$ is taken after 
TBA formalism is set up.  
  
The free energy of this model in the $L\goes\infty$ limit 
should be
\BE
{F\over L}=f_{\rm bulk}-{\theta\over L}\ln g +O(1/L^2 ) ,
\EE
where $g$ is the ground state degeneracy 
of the system with the boundary at $\sigma=0$. 
We focus on this $g$-function. 

The TBA procedure gives us an equation 
for the $g$-function in terms 
of the hole energy functions 
$\epsilon_r$ $(r=1,2,\cdots,\lambda+1)$\cite{fendley-saleur-warner}
\footnote{Strictly speaking, we are going to consider 
just the difference of the 
$g$-functions at two different temperatures.}:
\BE
\ln g=\sum_{r=1}^{\lambda+1}\int^{\infty}_{-\infty}{dv\over 2\pi}
\kappa_{r}(v -\ln(\theta/\theta_B ))\ln(1+e^{-\epsilon_r (v)}),
\label{eq:g-fn}
\EE
where $\kappa_r$ are 
the kernels whose 
Fourier transforms are 
\BEA
{\tilde\kappa}_n (y)&=&{\sinh y\over 2\sinh y\cosh\lambda y},\NN\\
{\tilde\kappa}_\lambda (y)&=&
   {\sinh(\lambda-1)y\over 2\sinh 2y\cosh\lambda y},\quad 
{\tilde\kappa}_{\lambda+1}(y)={\tilde\kappa}_\lambda (y)+{1\over 2\cosh y},\\
{\tilde\kappa}_r (y)&=&\int^{\infty}_{-\infty}{dv\over 2\pi}
e^{i2\lambda vy/\pi}\kappa (v)\NN .
\EEA
The hole energies $\epsilon_r (v)$ satisfy the following TBA equation 
\BE
\epsilon_r (v)=\sum_{s=1}^{\lambda+1}a_{rs}\int^{\infty}_{-\infty}
{dv'\over 2\pi}
{1\over\cosh(v-v')}\ln(1+e^{\epsilon_s (v')}), \label{eq:en1}
\EE
where $a_{rs}$ is the incidence matrix of $D_{\lambda+1}$-type 
Dynkin diagram;
\BEA
a_{ij}&=&\delta_{i,j+1}+\delta_{i,j-1}\qquad (i,j=1,2,\cdots,\lambda-1),\NN\\
a_{\lambda,j}&=&\delta_{j,\lambda-1},\quad 
a_{\lambda+1,j}=\delta_{j,\lambda-1}.
\EEA

In the two limits, $\theta/\theta_B =0$ and $\theta/\theta_B =\infty$, 
the above TBA equation can be solved 
analytically\cite{fendley-saleur-warner}. We call the former 
limit infrared (IR) and the latter ultraviolet (UV). 
The difference of the boundary entropies in these two limit is
\BE
{g_{\rm UV}\over g_{\rm IR}}={R\over 2\pi\sqrt{\ap}}.
\label{eq:g-ratio}
\EE
As is conjectured by the $g$-theorem\cite{affleck-ludwig}, 
$g$ decreases 
along the RG flow from UV to IR if $R>2\pi\sqrt{\ap}$, {\it i.e.} 
the boundary perturbation is relevant. 
$g$'s in the two limits have been identified 
with the respective values of the  
tachyon actions\cite{harvey-kutasov-Martinec,tensionisdimension}. 
We can compare the tensions 
of D25- and D24-branes, $\tau_{25}$ and $\tau_{24}$ respectively.  
In this view, we should set 
$g_{\rm UV}=\tau_{25}R$ and $g_{\rm IR}=\tau_{24}$. 
Thus, we get the well-known relation 
$\tau_{24}=2\pi\sqrt{\ap}\tau_{25}$. 

Even at general $\theta$, $g$ is obtained 
numerically by means of TBA.  
We expect that 
this $g$ will give the string action even in an intermediate region 
between the open string vacuum and the closed string one.  
The quantity $\ln(\theta/\theta_B )$ is identified with 
the RG scale.  
As an example, let us calculate the $g$-function for $\lambda=2$ 
({\it i.e.} $R=2\sqrt{3\ap}\pi$)  
case explicitly. The plot $\ln(\theta/\theta_B)$-$\ln g$ is shown in 
Figure 1. 
\begin{figure}
  \epsfxsize = 8 cm   
  \centerline{\epsfbox{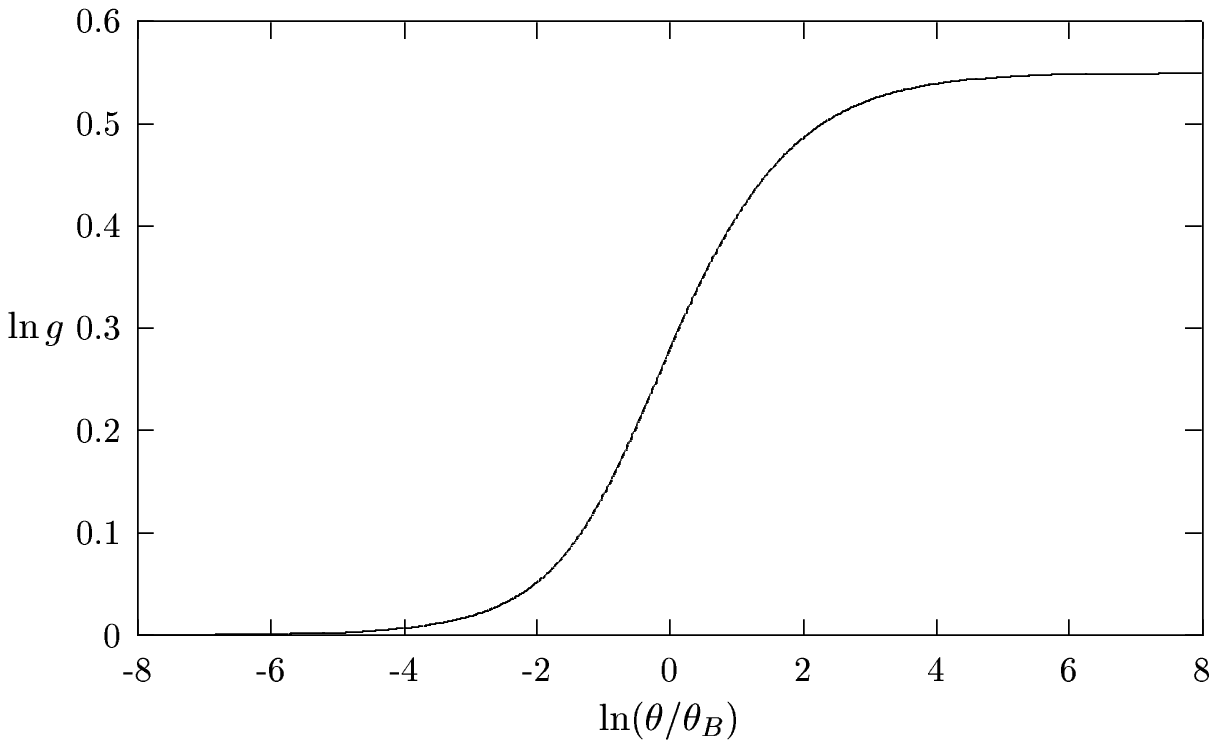}}
  \centerline{~}
  \centerline{{\tt Figure 1}: Boundary entropy for $R=2\sqrt{3\ap}\pi$}
\end{figure}
As is seen in Figure 1, $g$ becomes stationary both as    
$\ln(\theta/\theta_B)\goes -\infty$ and as 
$\ln(\theta/\theta_B)\goes +\infty$. Having this aspect of the 
$g$-function in mind, we would like to gain more insight 
into the RG behavior of the 
tachyon condensation. 

Let us consider the behavior of the $g$-function at 
large $\theta/\theta_B $ in field theory analysis
\cite{affleck-ludwig}. 
Let the dimension of the boundary term be $\Delta$. 
We see that 
the $\beta$-function of the coupling $\zeta$ is
\BE
\beta(\zeta)={d\zeta\over d\ln|x|}=(1-\Delta)\zeta +O(\zeta^2),
\EE
where $|x|$ is the inverse of the momentum cut-off at the 
boundary and equal to the ratio, $(\theta/\theta_B )^{-1}$.  
Thus, we see the relation 
$\zeta\sim (\theta/\theta_B )^{-(1-\Delta)}$ 
for large $\theta/\theta_B$. 
Upon taking ($\ref{eq:deS}$) into account, we obtain the 
asymptotic behavior 
\BE
g\sim g_{\rm UV}-c_0 (\theta/\theta_B )^{-2(1-\Delta)} \label{eq:afg}
\EE
for small $\zeta$, where $c_0$ is a constant. 
We will compare this with the tachyon action later. 
{\bf Tachyon field and its energy}

Let us consider the tachyon field with the co-dimension one 
{\it i.e.} the case in which 
the tachyon field depends on just one 
coordinate $X^{25}(=x)$. 
The action obtained in 
\cite{gerasimov-shatashvili, kutasov-marino-moore} is 
\BE
S=\tau_{25}V_{25}\int^{+\infty}_{-\infty}\left[ 
\ap e^{-T}{\dot T}^2 +V(T)
\right] dx, \label{eq:Sh-action}
\EE
where $V(T)=e^{-T}(T+1)$ is the tachyon potential, 
${\dot T}=dT/dx$, and 
$V_{25}$ is the volume of 25-dimensional space-time. 
Here we have ignored the higher derivative corrections.  
Let us set $\tau_{25}V_{25}$ to be $1/2\pi$ for simplicity and 
$\ap=1$.
Let us consider classical solutions of this 
action (\ref{eq:Sh-action}). The equation of motion obtained from  
(\ref{eq:Sh-action}) is 
\BE
2{\ddot T}-{\dot T}^2 -e^T V'(T)=0, 
\EE
which can be integrated once to give 
\BE
{\dot T}=\pm e^{T/2}\sqrt{E+V(T)}. \label{eq:emtachyon}
\EE
Here $E$ is a constant that can be regarded as energy. 
If $-1\leq E<0$,  
the tachyon field $T(x)$ looks like a classical lump. 
That is, $T(x)$  oscillates between $T_i$ and $T_f$. 
The constants $T_i$ and $T_f$ are respectively 
the negative and positive solutions of the equation 
\BE
E+V(T)=0 \label{eq:titf}
\EE
and we set $T(0)=T_i$ (see Figure 2). 
\begin{figure}
  \epsfxsize = 6 cm   
  \centerline{\epsfbox{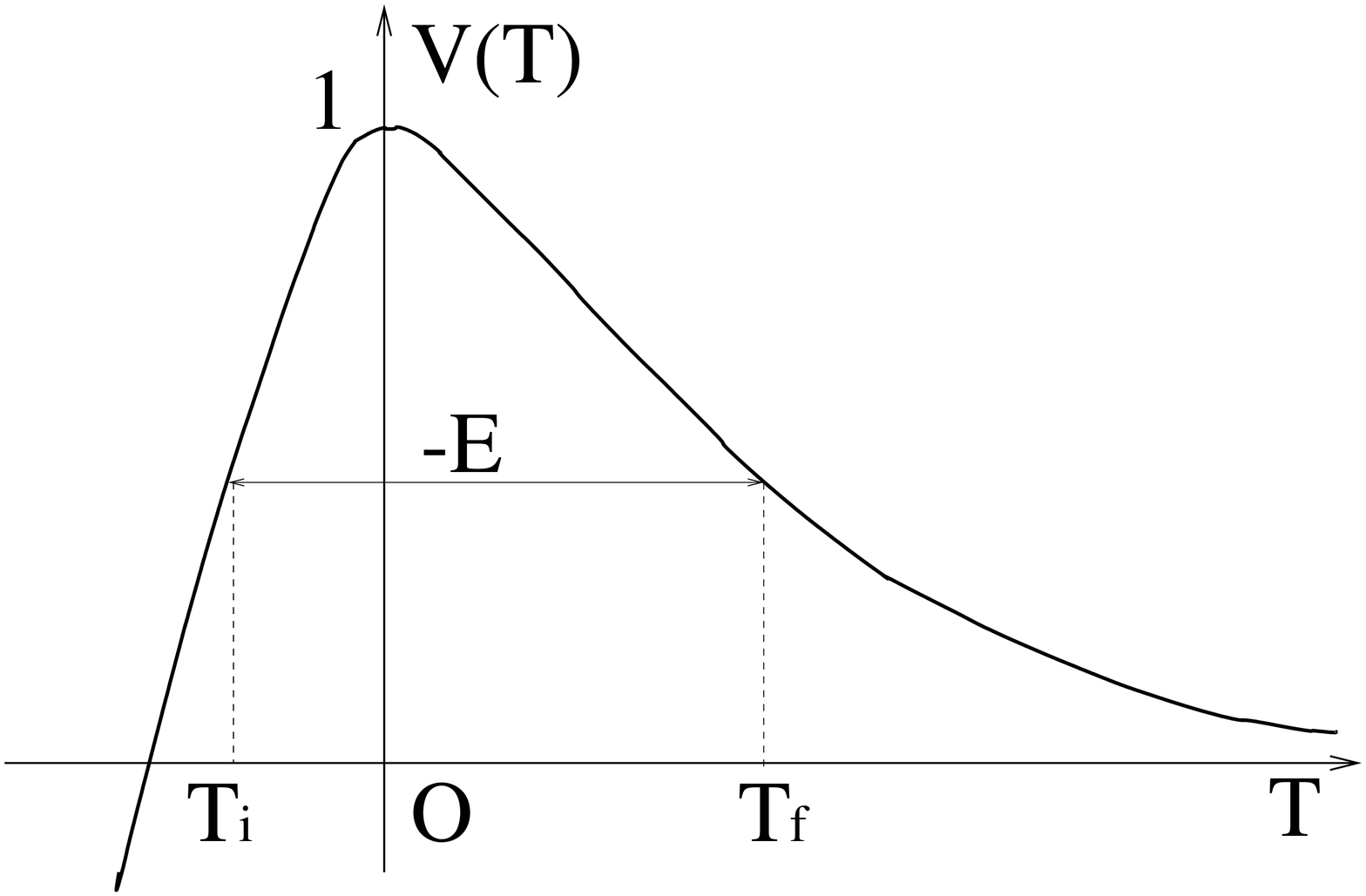}}
  \centerline{~}
  \centerline{{\tt Figure 2}: Classical solution}
\end{figure}
For example, setting $E=-1$, we see $T_i =T_f =0$, 
thus, $T(x)=0$ for all $x\in(-\infty,+\infty)$, which is on the 
UV fixed point and regarded as the 
tachyonic open string vacuum. Another example is $E=0$ which 
has $T_i =-1$ and $T_f =+\infty$. In this case, 
(\ref{eq:emtachyon}) is easily integrated to give  
\BE
T(x)=-1+x^2 /4.
\EE 
This form 
is already used in \cite{kutasov-marino-moore}. 

Let us evaluate (\ref{eq:Sh-action}) for given 
energy ($-1\leq E\leq 0$). Because (\ref{eq:Sh-action}) naively 
diverges, it is 
regularized by introducing the cut-off like 
\BE
S(E,R)=\int^{R/2}_{-R/2}{dx\over 2\pi}\left[ 
e^{-T}{\dot T}^2 +V(T)
\right] .
\EE
This cut-off $R$ is identified with that in (\ref{eq:action}). 
Using (\ref{eq:emtachyon}), we see
\BE
S(E,R)=-{RE\over 2\pi}+4\oint^{T_R }_{T_i}{dT\over 2\pi}\,e^{-T/2}
\sqrt{E+V(T)}, \label{eq:actionreg}
\EE
where we have set $T(R/2)=T(-R/2)=T_R $. From the form of  
(\ref{eq:actionreg}), we conclude $S(-1,R)=R/2\pi$ and 
$S(0,R)=e/\sqrt{\pi}+O(e^{-R^2}/R)$. We should note that the 
precise value of $S(0,\infty)$ should be 1 as is seen in 
(\ref{eq:g-ratio}). In order to get this precise value, a more 
careful treatment to include the higher derivative terms, 
which we have ignored in (\ref{eq:Sh-action}), 
is necessary\cite{kutasov-marino-moore}.   

Let us concentrate on the region near the UV fixed point, 
namely the region where  
the condensation is forming. Let 
$E=-1+\epsilon^2 \quad (0<\epsilon\ll 1)$. 
Because $|T_i|, |T_R|, T_f \ll 1$ there, we approximate 
$V(T)\cong 1-T^2 /2$. Thus, we see 
$-T_i \cong T_f \cong\sqrt{2}\epsilon$. 
After some elementary calculation, we obtain, for example, 
\BE
S(-1+\epsilon^2 ,R)=R/2\pi-\epsilon^2 \cdot\sqrt{2}\sin(R/\sqrt{2})
+O(\epsilon^3)\quad {\rm for}\quad 2\sqrt{2}\pi<R<3\sqrt{2}\pi.
\label{eq:UVaction}
\EE 
Let us compare the action (\ref{eq:UVaction}) with the $g$-function 
(\ref{eq:afg}) of BSG model with TBA.  
Comparing the scaling of $g$ with $\theta/\theta_B$ 
and the scaling of $S$ with $\epsilon$, we find 
\BE
\epsilon\propto (\theta/\theta_B )^{-(1-\Delta)}.
\EE
Then we expect $S(-1+\epsilon^2 ,R)=g(\theta/\theta_B ,R)$ after 
fixing the constant $c_0$ in (\ref{eq:afg}) appropriately. 
Let us consider the case where the cut-off is $R=2\sqrt{3}\pi$. 
We have already shown the flow of the $g$-function in Figure 1. 
The string action $S$ can also be calculated  
numerically by using (\ref{eq:titf}) and (\ref{eq:actionreg}).  
Because, in this case, the boundary interaction 
has the dimension, $\Delta=1/3$, 
the two scaling parameters $E$ and $\theta/\theta_B$ 
should be related as   
$E+1\sim (\theta/\theta_B )^{-4/3}$. 
The plots are shown in 
Figure 3, where we put $\epsilon\cong 5.4(\theta/\theta_B)^{-2/3}$. 
Figure 3 indicates the numerical agreement of the 
scalings for $g$ and $S$ for $\sqrt{E+1}\lsim 0.2$.  
\begin{figure}
  \epsfxsize = 8 cm   
  \centerline{\epsfbox{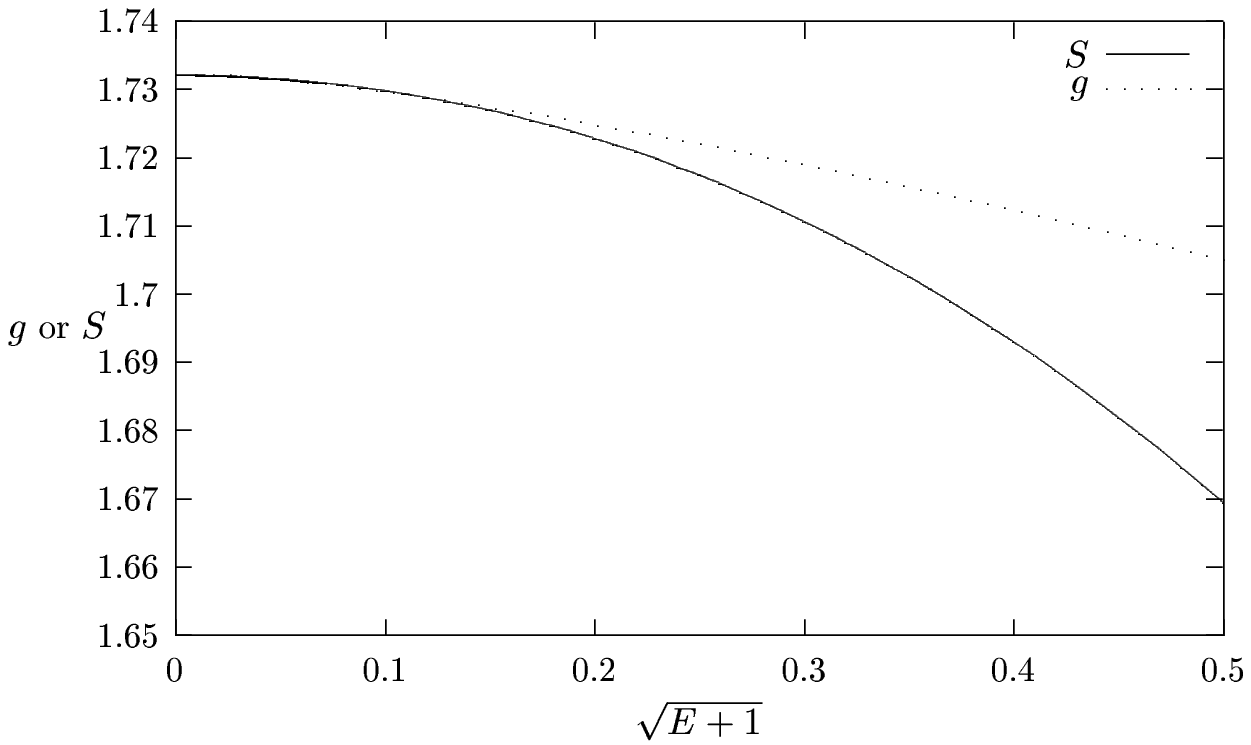}}
  \centerline{~}
  \centerline{{\tt Figure 3}: $g$-function and tachyon action for 
$R=2\sqrt{3}\pi$}
\end{figure}
When approaching the IR fixed point, 
$\theta/\theta_B \goes 0$ and $E\goes 0$, 
the higher derivative correction for (\ref{eq:Sh-action}) 
must become important in order that the relation between 
$S$ and $g$ holds in this region as well. \\
~\\
{\bf Acknowledgments}

The authors thank T. Nakatsu, N. Ohta, K. Okuyama and T. Suyama 
for helpful discussions. 


\end{document}